\documentclass[11pt,onecolumn,preprintnumbers]{revtex4}

\usepackage{amsmath,amsfonts,amssymb}
\usepackage[utf8]{inputenc}
\usepackage{graphicx}
\usepackage{epstopdf}
\usepackage{float,placeins}
\usepackage[caption=false]{subfig}
\usepackage[driverfallback=dvipdfm,hyperfigures,breaklinks,colorlinks,citecolor=blue,linkcolor=red]{hyperref} % hyperlink
\usepackage{color}

\begin{document}

\title{An improved Belief Propagation algorithm finds many Bethe states in the random field Ising model on random graphs}

\author{G. Perugini}
\affiliation{Dipartimento di Fisica, Universit\`a ``La Sapienza'', P.le A. Moro 5, I-00185, Rome, Italy}
\author{F. Ricci-Tersenghi}
\affiliation{Dipartimento di Fisica, Universit\`a ``La Sapienza'', P.le A. Moro 5, I-00185, Rome, Italy}
\affiliation{INFN-Sezione di Roma 1, and CNR-Nanotec, UOS di Roma.}

\begin{abstract}
We first present an empirical study of the Belief Propagation (BP) algorithm, when run on the random field Ising model defined on random regular graphs in the zero temperature limit.
We introduce the notion of extremal solutions for the BP equations and we use them to fix a fraction of spins in 
their ground state configuration. At the phase transition point the fraction of unconstrained spins percolates and 
their number diverges with the system size. This in turn makes the associated optimization problem highly non trivial in the critical region.
Using the bounds on the BP messages provided by the extremal solutions we design a new and very easy to implement 
BP scheme which is able to output a large number of stable fixed points. On one side this new algorithm is able
to provide the minimum energy configuration with high probability in a competitive time. 
On the other side we found that the number of fixed points of the BP algorithm grows with the system size in the critical region. 
This unexpected feature poses new relevant questions on the physics of this class of models.
\end{abstract}
\maketitle
\section{Introduction}
One of the main features of disordered systems is the existence of many thermodynamic states, eventually metastable states \cite{Mezard1987}.
This assumption is at the basis of the replica symmetry breaking (RSB) theory, which is indeed proven to be correct for disordered models defined on fully connected topology, as the Sherrington-Kirkpatrick (SK) model.
The same assumption has made possible to achieve a very rich and accurate description of the space of solutions in constraint satisfaction problems \cite{Mezard2009}.

However, to the best of our knowledge, there exist no algorithm which is able in general to find these many different states that we usually count via the replica or the cavity method. Obviously there are some easy cases where some of these states can be identified straightforwardly (e.g. ferromagnetic models or planted models \cite{Zdeborova2016}), but for a general disordered model we are not aware of any algorithm that would output at least some of the states that characterize the Gibbs probability distribution in a given sample.

The same definition of a pure state in a disordered model is a delicate issue: indeed the standard trick of imposing the right boundary conditions is not easy to implement, not only because the number of different states may be very large, but also because for a generic graph the boundary is not well defined (think e.g. in random graphs or fully connected graphs).

Given a model of $N$ interacting variables $\underline\sigma = \{\sigma_i\}_{i=1,\ldots,N}$, one would like to find the minimal decomposition of the Gibbs measure
\begin{equation}
\mu(\underline\sigma) = \sum_\alpha w_\alpha\, \mu^\alpha(\underline\sigma)\;,
\label{eq:decomp}
\end{equation}
such that the measure $\mu^\alpha(\underline\sigma)$ within state $\alpha$ has the clustering property, i.e. connected correlations decay fast enough at large distance, and thus distant variables are mostly independent (and the corresponding measure approximately factorizes).

In this work we consider models defined on sparse random graphs. These models can be solved exactly under the Bethe-Peierls approximation as long as correlations decay fast enough along the edges of the graph.
Moreover they show more realistic physical properties, with respect to models defined on fully-connected network: e.g.\ the latter have interaction couplings scaling with an inverse power of the the system size and critical lines in the field vs.\ temperature plane diverging in the $T\to0$ limit.

For models defined on sparse random graphs, it is natural to identify the measure within a state $\mu^\alpha(\underline\sigma)$ with a Bethe measure \cite{Mezard2009}, i.e.\ a measure that can be, at least locally, factorized as
\[
\mu^\alpha(\underline\sigma) \simeq \prod_{(ij) \in E} \frac{\mu^\alpha_{ij}(\sigma_i,\sigma_j)}{\mu^\alpha_i(\sigma_i)\mu^\alpha_j(\sigma_j)} \prod_{i\in V} \mu^\alpha_i(\sigma_i)\;,
\]
where the first product runs over all edges of the graph (we assume for simplicity that variables interact only pairwise, the generalization to higher order interactions being straightforward) and the second is over all vertices, while $\mu^\alpha_i$ and $\mu^\alpha_{ij}$ are marginal probabilities over single variables and pairs of neighboring variables respectively.

In principle each Bethe measure $\mu^\alpha(\underline\sigma)$ can be put in correspondence with a fixed point of the Belief Propagation (BP) algorithm \cite{Mezard2009}, however in practice we are not aware of any numerical protocol that outputs such BP fixed points in a generic disordered model.
The importance of these BP fixed points is also highlighted by recent results \cite{Coja2017} proving that any Gibbs measure on a random graph can be expressed as the superposition of a relatively small number of Bethe states, that can be put in correspondence to BP fixed points.

The aim of the present work is to introduce a new algorithm which is able to identify many different Bethe states $\mu^\alpha$ in the Random Field Ising Model (RFIM) defined on a random graph at $T=0$.
Our new algorithm is based on the usual BP algorithm \cite{Mezard2009}, which in the $T=0$ limit is also known as max-sum algorithm. BP is an iterative algorithm whose fixed points correspond to minima of the Bethe free energy \cite{Yedidia2005} and thus provides the local marginal probabilities $\mu^\alpha_{ij}$ and $\mu^\alpha_i$ within a Bethe state.
In the case of the RFIM it has been recently shown that the global minimum of the Bethe free energy at $T=0$ does actually correspond to the model ground state, irrespective of the graph which is defined on \cite{Chertkov2008}. 

While running the standard BP algorithm on a given sample of the RFIM one usually reaches 1 or at most 2 fixed points, without any guarantee of having found the one of lowest free energy; our new algorithm achieves many different fixed points and the probability that the ground state is among these many fixed points turns out to be practically 1.
Moreover, finding also the lowest excited states, our algorithm provides more physical information about the model than what can be extracted solely from the knowledge of the ground state \cite{Lucibello2014}.

\section{The model}
\label{sec:model}

The random filed Ising model is well known to the statistical physics community 
as one of the simplest disordered systems \cite{Nattermann1998}. It is defined by the following Hamiltonian
\begin{equation}
\label{ener}
H(\underline{s}) = - J \sum_{(i,j) \in E} s_i s_j - \sum_{i \in V} h_i s_i \,,
\end{equation}
where $E$ and $V$ are respectively the edge set and the vertex set of the interaction graph (e.g.\ a complete graph, a random graph, or a finite dimensional regular lattice). Variables $s_i=\pm1$ are $N=|V|$ Ising spins and we choose to work with random fields $\{h_i\}$ extracted from a Gaussian distribution of zero mean and unitary variance, that ensures to have a second order phase transition in the whole field vs.\ temperature plane, including on the $T=0$ axis.
%The coupling constant $J$ can be varied to study the different phase of the model. In order to have a well defined thermodynamic limit we need to have $J |E| = O(N)$: thus on a fully connected graph we scale $J \sim 1/N$, while, on a sparse random graph, where $|E| = O(N)$, we keep the coupling constant $N$-independent.

Varying $J$ the model undergoes a second order phase transition at $J_c$ between a paramagnetic phase for $J<J_c$ and a ferromagnetic phase for $J>J_c$.
Other kinds of long range order, e.g. a spin glass phase, dominating the thermodynamics have been excluded \cite{Krzakala2010, Chatterjee2015} for the model in Eq.~(\ref{ener}), that is the one with pairwise interactions between Ising spins (although replica symmetry breaking effects have been numerically detected in the $p$-spin version \cite{Matsuda2011}
and in the pairwise model with continuous spins \cite{Lupo2017}).
The absence of replica symmetry breaking (RSB) for the states dominating the thermodynamics does not imply the absence of many metastable states \cite{Krzakala2010}, which may affect dramatically the dynamical behavior in the out of equilibrium regime \cite{Ohr2017}.
Moreover RSB effects may be present exactly at the critical point where the ferromagnetic susceptibility diverges in the thermodynamical limit.

Renormalization group studies indicate that thermal fluctuations are subdominant and the model can be studied
directly at zero temperature \cite{Ogielski1986}. This has a great numerical advantage since a min-cut algorithm exists, that provides the ground state (GS) configuration in polynomial time \cite{Hartmann2002}.
It is worth noticing, that this min-cut algorithm provides a single GS, even in case of strong GS degeneracy: in other words it is not a good sampler of the Gibbs measure, neither it can give any information about the gap between the GS and the first excited state.

%\subsection{Identifying physical states in the fully-conncted RFIM}
%
%As a warm-up exercise we focus on the fully-connected version of the RFIM, which is defined by the energy function in Eq.~(\ref{ener}) with the replacement $J \to J/N$ in order to have a good thermodynamic limit.
%In the large $N$ limit any pure state $\alpha$ is identified by a set of magnetizations satisfying the naive mean-field equations, that at $T=0$ read
%\[
%m^\alpha_i = \text{sign}\left( h_i + \frac{J}{N} \sum_{j\neq i} m^\alpha_j \right)
%\]
%
%A configuration of the system is called 1-spin flip stable (1SFS) if each spin is aligned with its local field:
%$\Delta_i s_i > 0 , \forall i$. In order for a configuration $\{s_i\}$ to be a minimum of the energy it must be 1SFS.

%At low coupling strength $J$ the spins will tend to align to their random fields
%in order to minimize the energy, while when $J$ is large enough they will tend
%to align all together in same direction. At a coupling value $J_c(\sigma_h)$ a second
%order phase transition takes place from a paramagnetic phase to an ordered 
%ferromagnetic phase.

In this work we study the RFIM on random 4-regular graphs (RRG).
This ensemble consists of graphs where each vertex has exactly 4 neighbors randomly chosen (we avoid self-connections and double edges between a pair of nodes).
These graphs are locally tree-like, in the sense that the local neighborhood of a randomly chosen vertex converges with high probability to a tree in the $N \to \infty$ limit \cite{Mezard2009} and the typical size of loops is $\log(N)$.

In the last two decades a lot of attention has been devoted to the RFIM defined on random regular graphs.
Mainly the interest was focused on the non-equilibrium physics of the RFIM, as it emerged as a very effective model 
for Barkhausen noise and hysteresis in magnets (see \cite{Dhar1997} and references therein).
Much efforts over the years have been made to characterize analytically the out-of-equilibrium 
magnetization \cite{Dhar1997}, correlation functions \cite{Handford2012}, hysteresis loops \cite{Shukla2001}, 
and expansions toward fully connected models \cite{Illa2006}, just to mention few. 
The out-of-equilibrium Glauber dynamics too was solved in \cite{Ohta2010}.
The vast number of metastable non-equilibrium states was numerically analyzed in \cite{Rosinberg2008}, 
where different techniques were used to provide a clear description of the model complexity.

For what concerns the equilibrium physics of the model, beside the classic paper \cite{Bleher1998} where the bimodal random fields
version was studied, the only modern analytic study of the RFIM on a random regular graph is the one present in \cite{Morone2014}, 
where the model is studied at zero and finite temperature using the cavity method approach \cite{Mezard2001, Mezard2003}.
Moreover on such graphs the loops corrections can be handled analytically \cite{Ferrari2013,Lucibello2014a,Lucibello2014}.

\section{Belief propagation equations and extremal solutions} \label{sec:BP}

Belief Propagation (BP) is an iterative message-passing algorithm for solving the self-consistency equations that determines the minima of the Bethe free-energy \cite{Bethe1935}: these minima do actually corresponds to the physical states of the system and the one with lowest free-energy is the one dominating the thermodynamics.
It is known that the Bethe approximation is exact on trees \cite{Weiss1997, Weiss2000} 
as it relies on the independence of the neighborhood of a given spin when this spin is removed 
from the graph (hence the name cavity method in physics). 
When the graph has loops the width of the approximation made is generally unclear, 
though applying BP to graphs with loops is a consolidated practice, specially in computer science studies:
in this case is called loopy belief propagation.
On random graphs, which are locally tree-like, the failure of BP is directly related to the birth of long range correlations and thus phase transitions.

In this context the RFIM presents a lot of features that makes it a very interesting 
model to approach with the BP algorithm.
According to results in \cite{Krzakala2007}, since the model is replica symmetric, BP should provide the exact
thermodynamics in this case, a conjecture that was recently proven in \cite{Coja2016}.
However one must be aware of the fact that these conclusions only apply to the case where only one BP
fixed point exists, a condition that is far from being satisfied by most of the models that have been studied 
with the BP algorithm. And indeed we shall see that for the RFIM has more than one fixed point.
Nevertheless in \cite{Chertkov2008} it was shown that the zero temperature global minimum 
of the Bethe free energy coincide with the ground state of the problem, for every graph topology.
Strictly speaking, this means that among the (possibly many) fixed points of the BP algorithm there must be
the GS solution. 
This peculiar feature of the RFIM stimulated a lot of research tending to systematize BP as a provably exact
ground state solver for the problems with an energy function as the one in Eq.~(\ref{ener}), 
see for example \cite{Gamarnik2012}. A nice example with this taste is the tree-reweighted 
message passing scheme \cite{Kolmogorov2006, Kolmogorov2012}. 
A very interesting recent result is present in \cite{Tarlow2011} where it is demonstrated that with a proper calibration
BP can be made equivalent to the min-cut/max-flow algorithm, which is an exact solver.

All these results points out that indeed BP can be good even if loops are present. However to our knowledge
an empirical study of the real performance of BP on single instances is lacking.
The main advantage of a direct approach, as we will see, is to face the presence of many fixed points.
On one side this obviates most of our analytic understanding of BP. On the other side we shall see that 
the organization of these BP solutions seems to be relevant for the physics of the RFIM.

Given an energy function of the type (\ref{ener}) with pairwise interactions and binary variables,
the BP update rules reads (see \cite{Mezard2009}):
\begin{eqnarray} \label{eq:bpEqT}
h^{(t+1)}_{i \to j} & = & H_i + \sum_{k \in \partial i / j} u^{(t)}_{k \to i} \nonumber\\
u^{(t)}_{i \to j} & = & T \, \text{atanh} \left[
\tanh(J / T ) \tanh( h^{(t)}_{i \to j} / T)
\right]
\end{eqnarray}
for every edge $(i,j) \in E$. 
By taking the limit $T \to 0$ we obtain:
\begin{eqnarray} \label{eq:bpEq}
h^{(t+1)}_{i \to j} & = & H_i + \sum_{k \in \partial i / j} u^{(t)}_{k \to i} \nonumber\\
u^{(t)}_{i \to j} & = & \hat{u}_J(h^{(t)}_{i \to j})
\end{eqnarray}
where
\begin{equation} \label{eq:uhat}
\hat{u}_J(x) = 
\left\{
\begin{array}{rl}
-J & , \, x \leq J\\
x & , \, -J < x < J\\
J & , \, x \geq J
\end{array}
\right.
\end{equation}
These equations are also known as min-sum equations.
Once an initial value is assigned to the set of messages $\{u^{(0)}_{i\to j}\}$, 
we let them evolve with the update rule (\ref{eq:bpEq}) until a stopping criterion is met.
In this work we use the following prescription: if at a time $t^{*}$, 
$\max_{(i,j)\in E} |h^{(t^{*})}_{i \to j} - h^{(t^{*}-1)}_{i \to j}| < \epsilon$
for some accuracy $\epsilon$, then the messages have converged to a fixed point.
We found that such a procedure is highly stable and we used $\epsilon = 10^{-10}$.
Once the equations have converged, one can easily compute the spins value with the fixed point messages
via
\begin{equation} \label{eq:bpSpin}
\sigma_i = \text{sign}\left\{H_i + \sum_{k \in \partial i} u^{(t^{*})}_{k \to i} \right\},
\end{equation}
from which physical observable can be computed.

Given that the RFIM at zero temperature displays a second order phase transition
at a critical value of the coupling strength $J_c=0.54404(5)$,
one expects to find one stable fixed point in the paramagnetic region $J < J_c$, 
and two stable fixed points in the ferromagnetic regime $J > J_c$, where the 
one with the magnetization of the same sign of the sum of the random fields will be
the dominating one, i.e. the one with lower energy \cite{Lucibello2014} .
As we will see this picture is over-simplified, and turns out that 
in a broad region near the critical point there is a large
number of stable fixed points, depending on the disorder realization.

\begin{figure}[t] 
\begin{center}
\includegraphics[height=6cm]{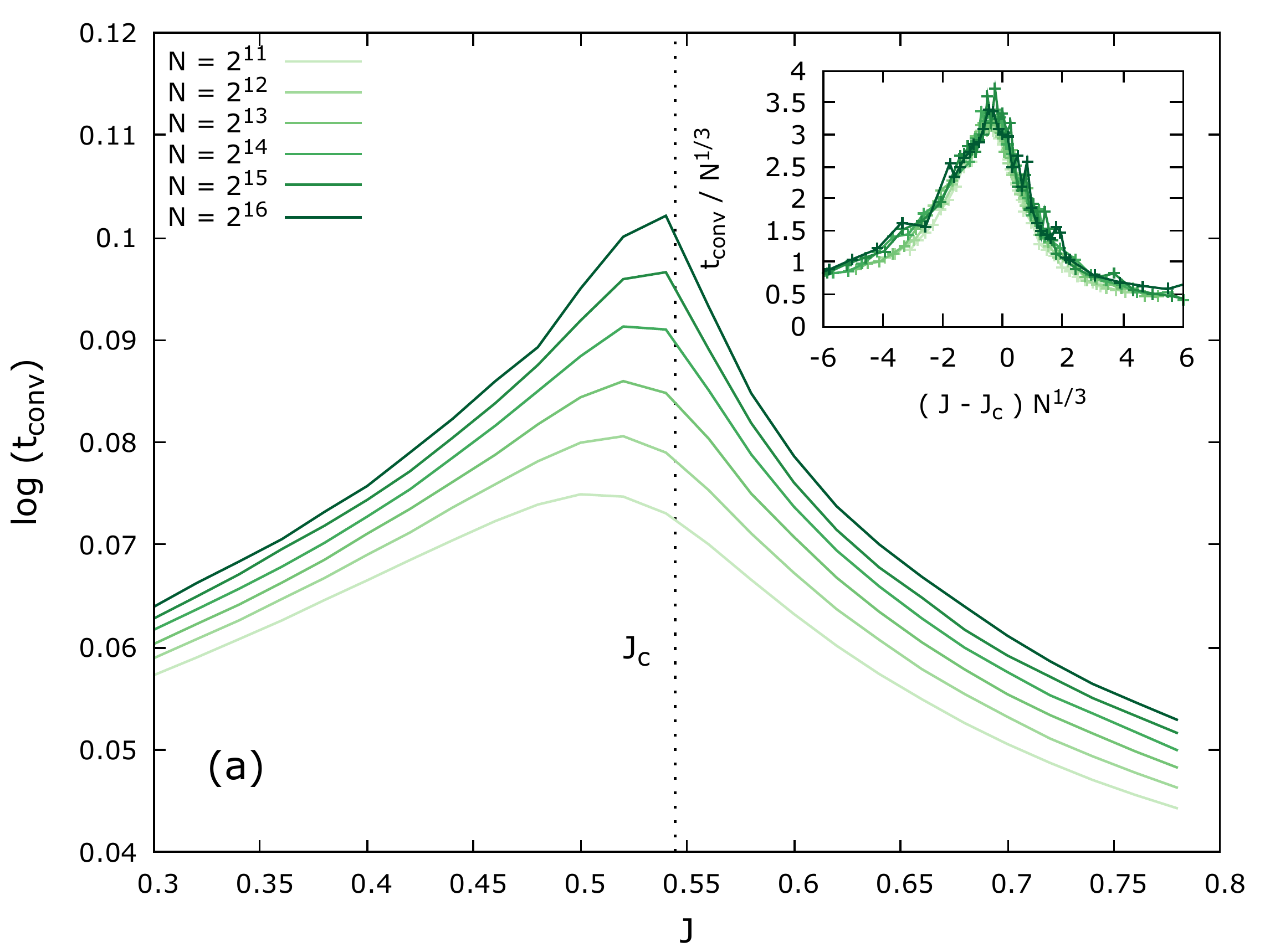}
\includegraphics[height=6cm]{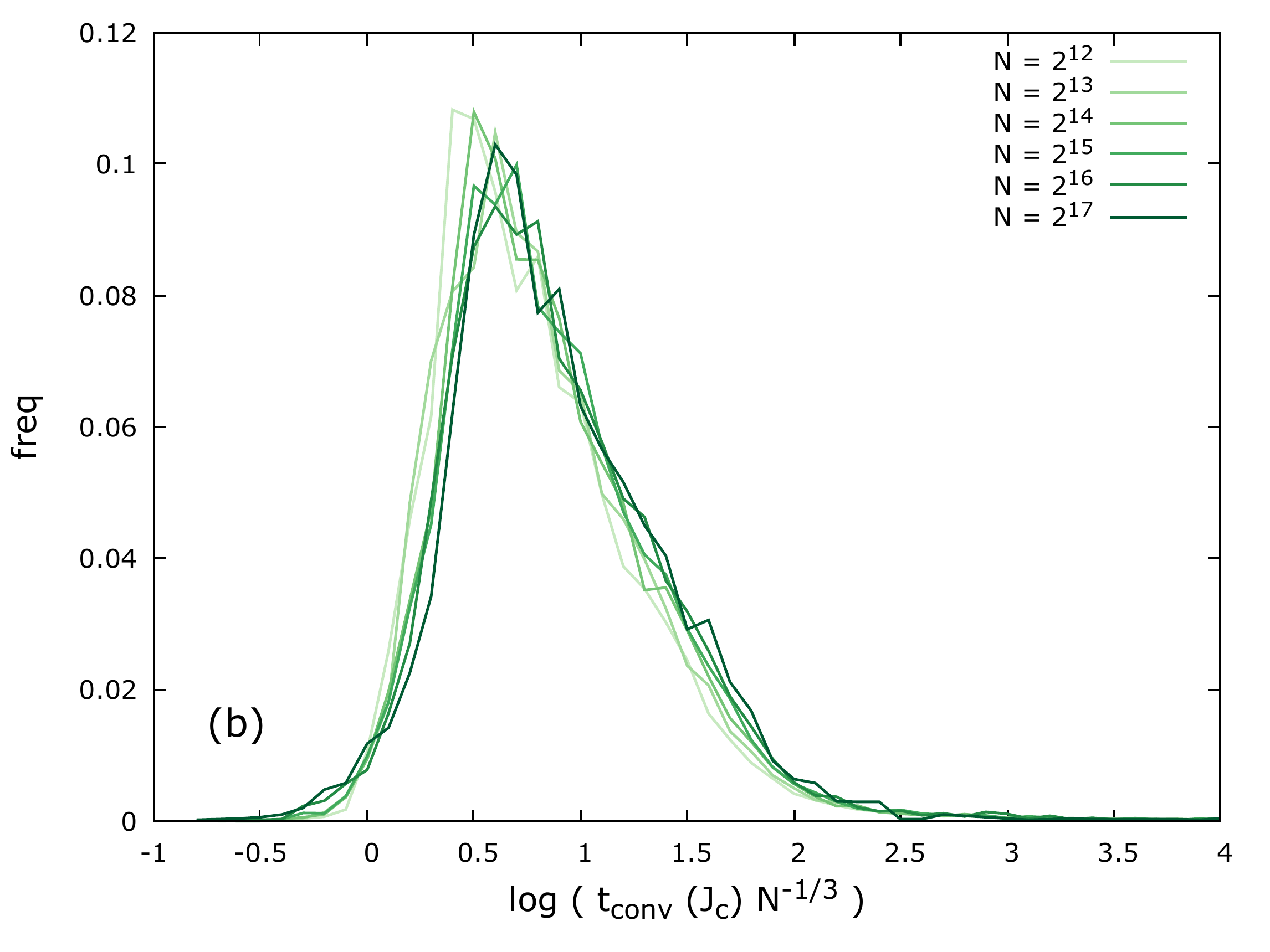}
\end{center}
\caption{(a) Average of the logarithm of the number of BP steps to converge to a fixed point (with a convergence criterion $\epsilon=10^{-10}$) as a function of the coupling strength $J$. In these plots we report the results for the $(+)$ IC only, since other IC provide similar results.
The maximum of $t_{\text{conv}}$ is found in the critical region, as expected. We used a sequential update with no damping. 
However we verified that the shape of the function $t_{\text{conv}} (J)$ is not qualitatively affected by the
particular choice of the update rule. In the inset we show the scaling with the system size $N$
of the convergence time in the critical region.
(b) The entire probability distribution of rescaled convergence time at the critical point $\log(t_\text{conv}(J_c)/N^{1/3})$ is almost size independent (and the same holds for other values of $J$).}
\label{fig:tconv}
\end{figure} 

From physical considerations, a natural choice for the initial condition (IC) of the BP messages 
is to bias them towards the sign of the sum of local random fields. Specifically we start with the two IC:
\begin{equation} \label{eq:icUpDown}
u^{(t=0), (\pm)}_{i \to j} = \pm J \, \text{sign} \left( \sum_{k = 1}^N H_k \right) \qquad \forall (i,j) \in E.
\end{equation}
In this way the $(+)$ IC is equivalent to set all the messages equal to their maximum value with the same sign
of the sum of the random fields of the specific instance, and analogously for the $(-)$ IC.
This choice is motivated from the following observation: when $J$ is small we expect that 
all the spins will prefer to align to their local random field, so that only one fixed point exists 
with zero average magnetization. In this situation the particular choice of the IC should play no role. 
On the other side, when the coupling constant is large enough
the spins will tend to align all together in the same direction, irrespective of the random fields, so 
we expect the $(+)$ IC to converge on the ferromagnetic minimum with lower energy and the $(-)$ IC to 
converge on the sub dominating one. As we will see in a while this intuition turns out to be true only away
from the critical region.

As the model is replica symmetric, for all $J$ we expect the BP algorithm to converge for every IC.
In Fig.~\ref{fig:tconv} we report the number of BP steps (where a single step corresponds to the update of all the $2|E|$ messages) to reach the convergence criterion, $t_\text{conv}$, as a function of the interaction
strength for the $(+)$ IC and we tested that the conclusion is the same for other choices for the IC.
As expected, the BP dynamics gets slower in the critical region and the convergence time seems to diverge at the critical point as $t_{\text{conv}} \sim N^{1/3}$.
In this respect the BP algorithm is competitive with the latest versions of the min-cut 
algorithm \cite{Boykov2004}, at least on locally tree-like graphs.
% serve citazione

However the BP algorithm is not guaranteed to converge on the minimum energy configuration (the GS) 
of a given graph. Although we know that the GS of the model should be a fixed point of the BP algorithm
\citep{Chertkov2008} it is not obvious at all how to initialize the messages in such a way to converge on the GS.
Moreover we have no way of asserting that a fixed point configuration is the one with the minimum possible energy,
although in the next section we shall give a criterion valid at least in the paramagnetic phase.

\begin{figure}[t]
\begin{center}
\includegraphics[height=6cm] {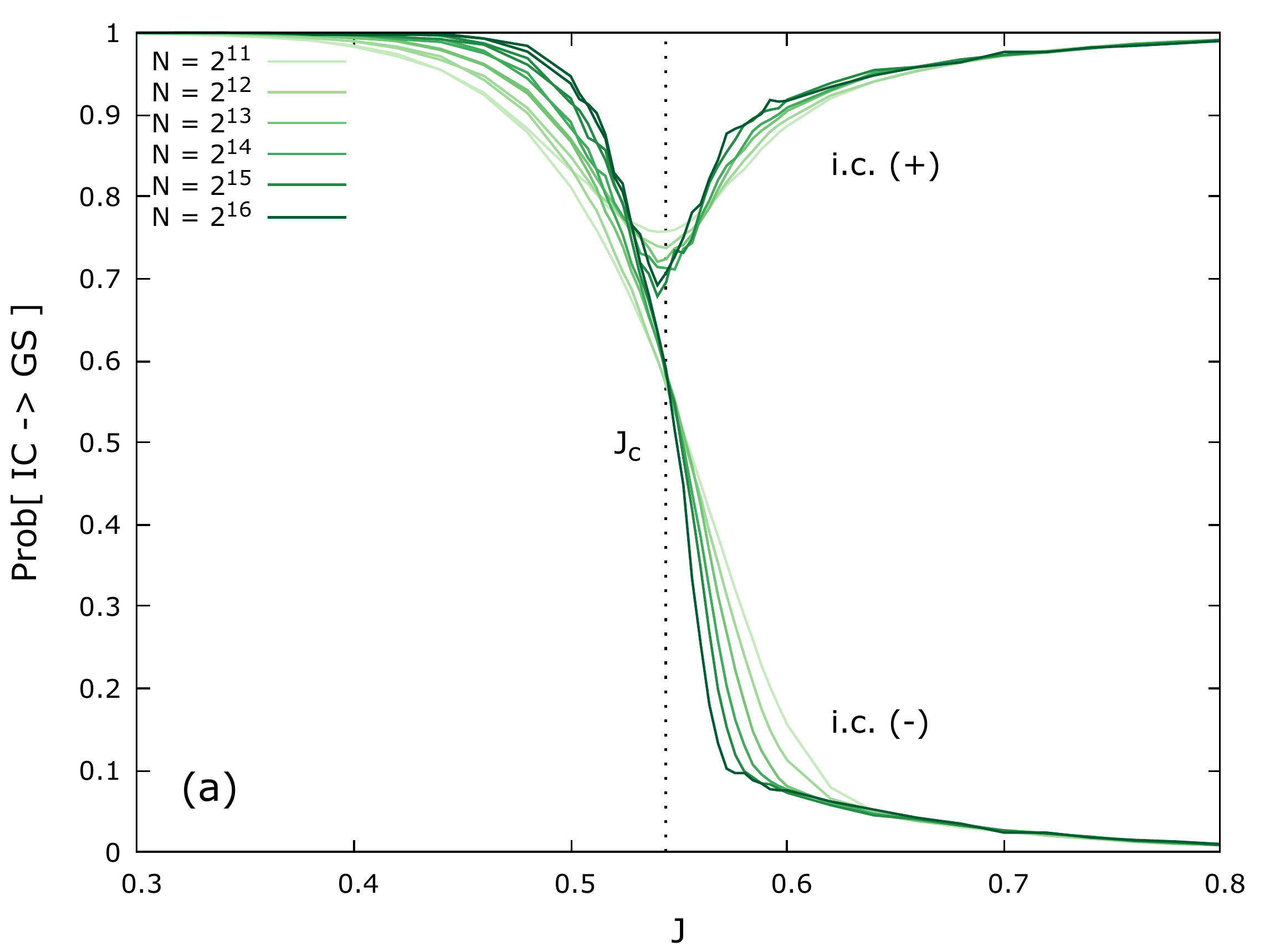}
\includegraphics[height=6cm] {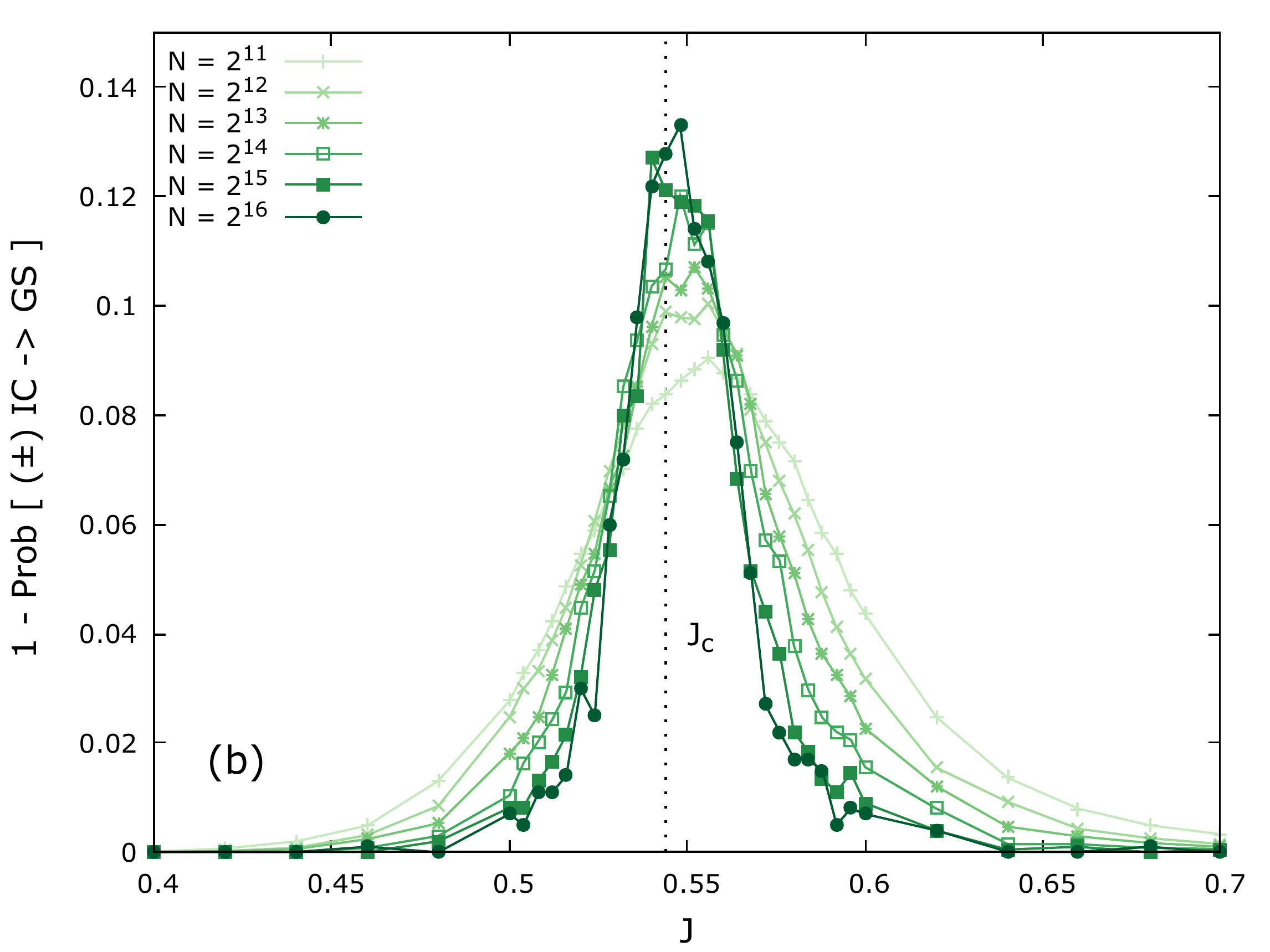}
\end{center}
\caption{(a) Probability that BP initialized with the $(\pm)$ IC --- see Eq.~(\ref{eq:icUpDown}) --- converges
to the GS, obtained with the min-cut algorithm. In the paramagnetic phase ($J < J_c$) there is just one fixed point with high probability and thus both IC converge on it with a probability tending to 1 in the large size limit. In the ferromagnetic phase ($J > J_c$) the two different IC converges to different free-energy minima, with the $(+)$ IC converging to the GS with a large probability, but still strictly smaller than 1.
(b) Probability than none of the $(\pm)$ IC converges to the GS is non-zero in the critical region, and increasing with size right at $J=J_c$.}
 \label{fig:pUpDown}
\end{figure} 

In order to understand the behavior of BP under the $(\pm)$ IC we measured the probability that it converges on the
GS configuration of the problem, obtained with the min-cut algorithm. 
The results are reported in Fig.~\ref{fig:pUpDown} for graphs of sizes ranging from $2^{11}$ to $2^{16}$. The
results are averaged over a number of realization for the disorder (graphs and random fields) that goes from
$5 \cdot 10^{4}$ for the smallest size to $5 \cdot 10^{3}$ for the biggest one.
As expected, the probability that the $(+)$ IC converges to the GS tends to one in the paramagnetic 
phase ($J < J_c$). In the ferromagnetic phase we found that there is a small but finite probability 
that the dominating minimum is not reached with the $(+)$ IC. This happens when the two IC converge
on two configurations with the magnetization of the same sign, where the one with the smallest modulus 
is reached with the $(-)$ IC and has the lower energy. Beside this phenomenon we can conclude that as far as
we are away from the critical region the two IC $(\pm)$ guarantee the convergence on the lowest 
energy configuration.

The situation drastically changes in the neighborhood of $J_c$. As the reader can see in the right panel of Fig.~\ref{fig:pUpDown} the probability that none of the $(\pm)$ IC converges on the GS is non-zero and grows with $N$ at $J_c$.
As we will see in more detail in Section \ref{sec:explor}, in the critical region many solutions of the BP equations 
appear and a level crossing phenomenon is at work, such that the GS can not be obtained by starting from an IC that corresponds to the GS at a nearby value of $J$.
Beside posing some question marks on the real nature of the phase transition, this picture makes the optimization problem very hard to solve with the standard BP algorithm. Although we know that the iterative BP equations must converge for every initial condition, we do not have any suggestion on how to initialize the BP messages in order to converge to the GS with high probability very close to $J_c$. We tested that the remaining trivial initial condition, i.e. the one with all the messages equal to zero, does not increases significantly the probability of success as it almost always converge on one of the two $(\pm)$ fixed points. The same holds for random initializations that achieve the GS with a smaller probability that starting from $(\pm)$ IC.

\section{Percolation of frozen variables} \label{sec:perc}

In this section we prove some properties of the extremal solutions, i.e.\ the fixed point solutions
obtained with the $(\pm)$ IC defined in Eq.~(\ref{eq:icUpDown}).
These results are preliminary to the definition of the algorithm that finds many different BP fixed points and rely on the following no-passing rule (NPR), first introduced by Middleton \cite{Middleton1992} in the context of charge density waves 
and later extended to the Glauber dynamics \cite{Dhar1997} and to the GS evolution \cite{Liu2007} in the zero temperature RFIM.

Let us adopt the convention that two vectors $\underline{v}$ and $\underline{v}'$ are partially ordered (to be indicated by $\underline{v} \succcurlyeq \underline{v}'$) if all their components satisfy $v_i \ge v'_i$.
Then, given two partially ordered initial configurations, $\underline{s}^{(\text{A})}(t=0) \succcurlyeq \underline{s}^{(\text{B})}(t=0)$, the NPR states that if they are evolved under ordered uniform fields satisfying $H^{(\text{A})}(t) \geq H^{(\text{B})}(t)$ for all times $t\ge0$, then the partial order among configurations is preserved for all times $t\ge0$.  
The validity of the NPR for the min-sum equations in the RFIM strictly follows from the definition
of the update rules, see eqs. (\ref{eq:bpEq}) and (\ref{eq:uhat}): thanks to the ferromagnetic couplings, each new message is a non-decreasing function of the old messages. Thus, if the initial messages satisfy
\begin{equation} \label{eq:messOrder}
u^{(t=0),\text{(A)}}_{i \to j} \geq u^{(t=0),\text{(B)}}_{i \to j}\, ,\qquad \forall (i,j) \in E,
\end{equation} 
then the same order must hold between messages at any time $t>0$.
Moreover when the fixed point of BP is reached, every spin will be computed as a non-decreasing function of the fixed points messages, such that the spin configurations will satisfy $\underline{s}^{\text{(A)}} \succcurlyeq \underline{s}^{\text{(B)}}$.

From Eq.~(\ref{eq:bpEq}) it is immediate to derive the following bound on the BP messages
\begin{equation} \label{eq:bpBound2}
-J \leq u^{(t)}_{i \to j} \leq J\;.
\end{equation}
The $(\pm)$ initial conditions, see Eq.~(\ref{eq:icUpDown}), do correspond to set the BP messages all equal and taking the largest (positive or negative) value allowed by the above bound.
This in turn implies that at any time on every edge of the graph a BP message cannot assume 
a value greater/lower than the value of the corresponding $(+)/(-)$ message
\begin{equation} \label{eq:bpBound3}
u^{(t),(-)}_{i \to j} \leq u^{(t),(*)}_{i \to j} \leq u^{(t),(+)}_{i \to j} \qquad
\forall t \,,\, \forall (i,j) \in E , 
\end{equation}
for every initial condition $(*)$.
Moreover the two initial condition $(\pm)$
will converge on the fixed points whose configurations are the one 
with the lowest and highest magnetization, as this is due to the NPR.
For this reason we shall call them the \emph{extremal solutions}.

Thanks to the inequality in Eq.~(\ref{eq:bpBound3}), if the fixed point messages $u^{(t^*),(-)}_{i \to j}$ and $u^{(t^*),(+)}_{i \to j}$ do coincide, then such a BP message is conserved in any other BP fixed point.
The same is true for the spin configurations obtained from the BP fixed point: if
\begin{equation} \label{eq:frozenSpin}
s^{(+)}_i = s^{(-)}_i \, , \qquad \text{where} \qquad 
s^{(\pm)}_i = \text{sign}\left\{H_i + \sum_{k \in \partial i} u^{(t^*),(\pm) }_{k \to i} \right\}.
\end{equation}
then spin $s_i$ must take the same value in all BP fixed points (i.e.\ in all the free-energy minima, including the GS) and we call it a \emph{frozen} spin.

Thanks to this simple property we can claim to have found the GS in case the two extremal solutions coincide (and this happen often in the paramagnetic phase).
Otherwise if only a finite fraction of the spins is frozen we can still reduce the complexity of the problem 
by removing these variables from the set of variables to be optimized over (the frozen spins actually change the field on the remaining variables, thus producing an effective RFIM of smaller size).
In general we expect the mean fraction of frozen spins to decrease with the coupling $J$: indeed for $J\ll J_c$ a unique fixed point exists and the extremal solutions do coincide, while for $J \gg J_c$ the extremal solutions do have very different magnetizations and practically no spin in common.

\begin{figure}{}
\begin{center}
\includegraphics[height=8cm]{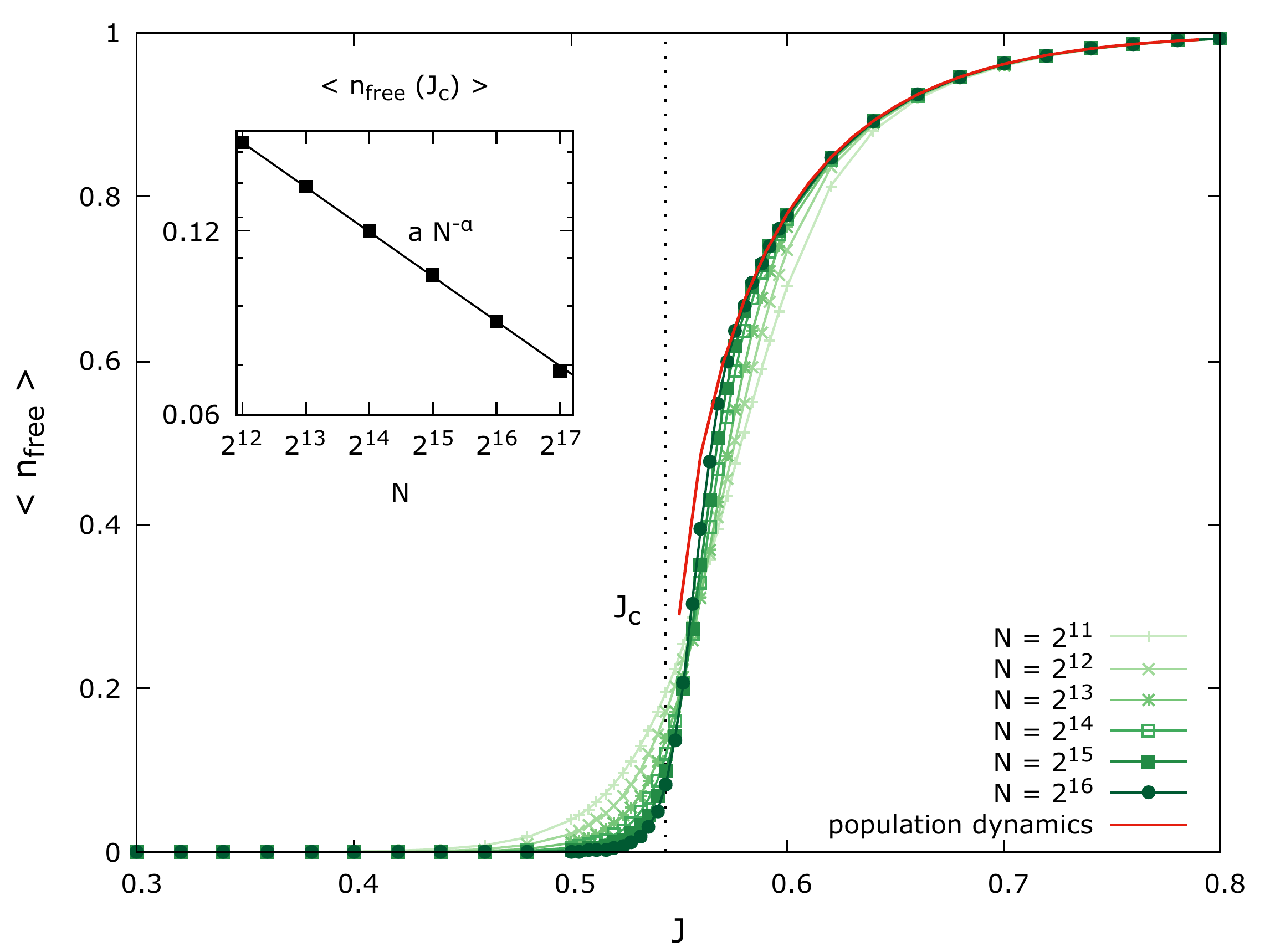}
\end{center}
\caption{Fraction of free spin variables, defined in Eq.~(\ref{eq:nFree}), for various system sizes as a function
of the interaction strength $J$. The data strongly suggest a percolation transition for free variables at $J_c$. For $J>J_c$ the data approach closely the infinite volume thermodynamic magnetization drawn with a red full line.
At $J_c$ the fraction of free spins decays as $n_{\text{free}} \sim N^{-\alpha}$ with $\alpha = 0.245 \pm 0.002$ (see the inset).}
\label{fig:nFree}
\end{figure}

Let us define the average fraction of free spins (i.e.\ non-frozen spins) as
\begin{equation} \label{eq:nFree}
n_{\text{free}}^{(N)} (J) \equiv \mathbb{E} \left[ \sum_{i=1}^N \frac{1 - s^{(+)}_i s^{(-)}_i}{2N} \right]\,.
\end{equation}
The mean fraction of free spins is shown in Fig.~\ref{fig:nFree} for a random
4-regular graph.

The data in Fig.~\ref{fig:nFree} strongly suggest the presence of a phase transition in $n_\text{free}$ exactly at the critical point $J_c$.
Moreover for $J>J_c$ the data seem to approach in the large size limit the curve for the absolute value of the thermodynamic magnetization $|m|$ computed via the population dynamics algorithm \cite{Mezard2009} (reported with a full red curve).
This can be explained by assuming that the extremal solutions are made as follows: a fraction $1-|m|$ of spins are shared by the two extremal solutions (this is the fraction of frozen spins) and give almost no contribution to the global magnetization (these spins are mostly aligned along the random field), while a fraction $|m|$ of spins are fully aligned, thus giving magnetizations $|m|$ and $-|m|$ to the two extremal solutions.

So, the ferromagnetic phase transition at $J_c$ seems to be related the the percolation of free variables.
At $J_c$ a single connected cluster of free spins starts spanning the entire graph and this makes the optimization problem of finding the GS much harder. Indeed at the critical point the fraction of free spins goes to zero as $n_{\text{free}}^{(N)} (J_c) \sim N^{-\alpha}$ with $\alpha \simeq 1/4$ (see inset in Fig.~\ref{fig:nFree}). This means that the mean number of free spins diverges at the phase transition as $N^{1-\alpha}$ giving us some insight on the nature of the difficulties that BP faces in finding the minimum energy fixed point among the many fixed points that appear in the critical region.

\section{Exploration of the Bethe free energy landscape} \label{sec:explor}

We present here a heuristic modification of the BP algorithm that is able to find several new BP fixed points (i.e.\ Bethe free energy minima) by relying on BP fixed points already found.
We always start with the two extremal IC $(\pm)$ defined in Eq.~(\ref{eq:icUpDown}), that converge to extremal solutions.
If the extremal solutions do coincide, then they provide the GS, and BP has no other fixed
points: thus the algorithm stops.
On the contrary, if extremal solutions are different, the frozen spins and the corresponding messages are fixed for the rest of the run, and the algorithm proceeds looking for more BP fixed points.
We saw in Section \ref{sec:BP} (see right panel in Fig.~\ref{fig:pUpDown}) that in the critical region 
the GS does not always correspond to a extremal solution, so it worth continuing the search for the possible GS.

The key question is how to initialize the non-frozen BP messages in order to find new BP fixed points (and hopefully the GS) without wasting time in random initializations.
Since the fixed point already found may have a rather large basin of attraction under the BP iteration, a reasonable initialization that is more likely to flow to a different fixed point (if any) is the following:
\begin{equation}
\label{eq:initBetween}
u^{(t=0)}_{i \to j} = \frac{1}{2} u^{(t^*),(+)}_{i \to j} + \frac{1}{2} u^{(t^*),(-)}_{i \to j}.
\end{equation}
In this way if $u^{(+), (t^*)}_{i \to j} = u^{(-), (t^*)}_{i \to j}$ we just initialize the BP message with the fixed point value.
On the contrary, the message is initialized as far as possible from the extremal messages, but also biased in their direction in case the extremal messages are correlated.
The idea is that if a new fixed point, say the $(0)$ fixed point, is found, then we can repeat the procedure by searching between $(+,0)$ and $(-,0)$ as before.
In this way we explore the space of BP messages, in the search for BP fixed points, by starting from a not too large and meaningful subset of IC.

In a nutshell our algorithm works as follows.
It keeps a list of BP fixed points (FP) that initially contains only the two extremal FP, $(+)$ and $(-)$.
For each pair of FP in this list, called them $(l)$ and $(r)$, the algorithm searches for new FP by starting from several IC belonging to the line $(l,r)$ joining the two FP.
The first search is performed from the IC $(c)$ being at the middle of segment $(l,r)$.
If BP initialized in $(c)$ converges to one of the two parent FP, say the $(r)$ FP, then the bounds in Eq.~(\ref{eq:bpBound3}) allow us to exclude the segment $(c,r)$ in the search for new and different FP; in this case the search continues with IC in the middle of segment $(l,c)$.
On the contrary, if BP initialized in $(c)$ converges to a new FP, this new FP is added to the list of FP and the search continues with IC both in the middle of $(l,c)$ and in the middle of $(c,r)$.
Each segment is analyzed (i.e.\ used to produce new IC) as long as it is larger than a given minimal length $\Delta$ (we have used $\Delta=2^{-32}$, but we have checked our results being largely independent from this minimal length).

The running time of the algorithm depends on several factors: the time required by BP to converge, that grows at most as $N^{4/3}$ in the critical region (being linear in $N$ far from the critical point); the number of pairs of FP used to generate the IC, $N_\text{sol}^2$, being $N_\text{sol}$ the number of FP found; a factor $\log(1/\Delta)$ proportional to the typical number of calls to the BP algorithm per segment analyzed.
The total time complexity of the our algorithm scales with the system size at most as $\mathcal{O}(N^{4/3})$, being the factor $\log(1/\Delta)$ size-independent and the mean number of solutions to BP equations $\langle N_{\text{sol}}\rangle$ a very slowly increasing function of $N$ (see discussion below).
Computationally this is slightly more expensive than running an exact solver such as min-cut, but it 
outputs a large number of solutions that can provide much more information on the physics of the RFIM.

\begin{figure}[t] 
\begin{center}
\includegraphics[height=8cm] {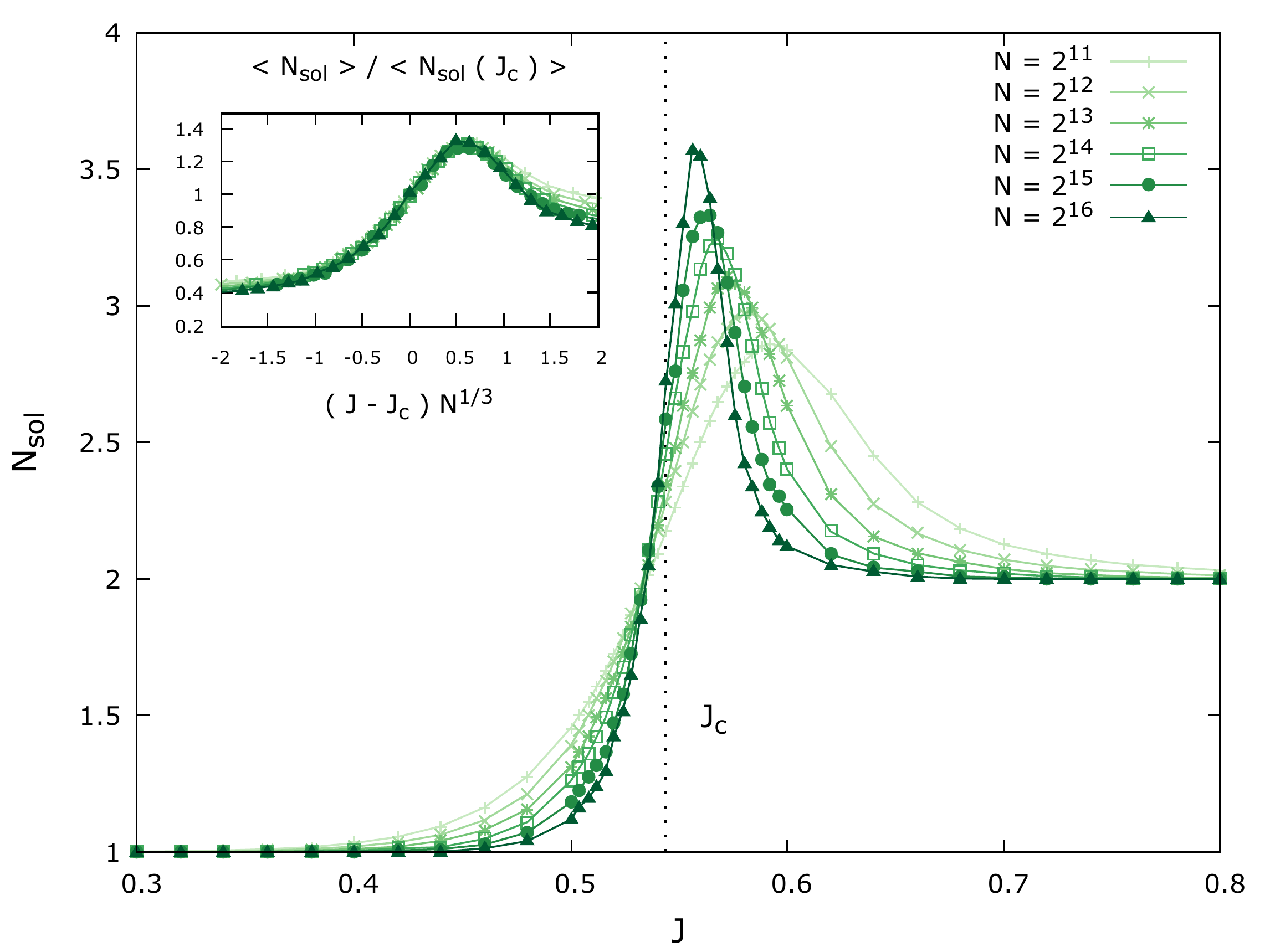}
\end{center}
\caption{Mean number of BP solutions (i.e.\ free-energy minima) found by our improved BP algorithm as a function of $J$ for various system sizes.
As expected the number of solutions tends to 1 in the low coupling regime, and to 2 in the ferromagnetic phase. In the critical region, however, the mean number of solutions increases with
the system size. (Inset) Scaling of the mean number of solutions in the critical region.}
\label{fig:nSolRRG}
\end{figure} 

The mean number of BP fixed points (i.e.\ solutions to BP equations) $\langle N_\text{sol} \rangle$ that our algorithm outputs is reported in Fig. \ref{fig:nSolRRG}.
Far from the critical point, for $|J - J_c| \gg 1$, typically 
the algorithm does not provide additional information than running BP with
the extremal initial conditions (see Fig.~\ref{fig:pUpDown}).	
In the paramagnetic phase the extremal solutions coincide and the bisection function
does not need to be called at all, since no other fixed point is admitted.
In the ferromagnetic phase with high probability only the segment joining the $(\pm)$
fixed points needs to be explored and no other solution is usually found.

In the critical region the number of BP fixed point (i.e. free-energy minima or Bethe states) found by our improved BP algorithm grows slowly with $N$. Such data can be well collapsed (see inset in Fig.~\ref{fig:nSolRRG}) by scaling it vertically according to the mean number of BP states at criticality, $\langle N_\text{sol}(J_c)\rangle$, and horizontally using the standard scaling variable $N^{1/3}(J-J_c)$.

\begin{table}[t]
\begin{center}
\begin{tabular}{|p{1cm}|c|c|}
\hline
\centering     $N$    &  \# samples    &  $1 - P_{\text{success}} (J_c)$ \\
\hline
\centering $2^{17}$ & $5\cdot10^3$ & $(2\pm2)\cdot10^{-4}$ \\
\hline
\centering $2^{16}$ & $1\cdot10^4$ & $(1\pm1)\cdot10^{-4}$ \\
\hline
\centering $2^{15}$ & $2\cdot10^4$ & $(0\pm3)\cdot10^{-5}$ \\
\hline
\centering $2^{14}$ & $4\cdot10^4$ & $(5\pm4)\cdot10^{-5}$ \\
\hline
\centering $2^{13}$ & $8\cdot10^4$ & $(5\pm3)\cdot10^{-5}$ \\
\hline
\centering $2^{12}$ & $16\cdot10^4$ & $(3\pm1)\cdot10^{-5}$ \\
\hline
\end{tabular}
\end{center}
\caption{Probability that our modified BP algorithm does not find the GS configuration at the critical point $J_c$ for various system sizes.
These data are well fitted by an $N$-independent probability of failure $1 - P_{\text{success}} \simeq 3\cdot10^{-5}$.}
\label{tab:probGS}
\end{table}

The presence of an increasing number of free-energy minima in the critical region, makes the search for the ground state a non-trivial problem here and explains why the extremal solutions often differ from the GS in this region (see right panel in Fig.~\ref{fig:pUpDown}).
Nevertheless, even at $J_c$, where the problem of finding the GS is most difficult, our improved BP algorithm misses the true GS only in a really tiny fraction of samples: in Table~\ref{tab:probGS}  we report such a number of samples, that are statistically compatible with a probability of missing the true GS independent of $N$ and close to $3\cdot10^{-5}$.
In the very rare samples where our algorithm does not find the true GS, the lowest free-energy minimum found lays above the true GS by $\mathcal{O}(10^{-4})$. Moreover we believe that with a proper calibration 
of the algorithm (e.g.\ by using some appropriate damping) the probability of finding the GS can be made even higher.
Here, however, our primary interest is on exploring efficiently the Bethe free-energy minima 
and so we are not going deeper with the possible use of our improved BP algorithm for solving the associated 
optimization problem, thou we believe that this could be a promising direction to follow.

At this point a comment is mandatory. It is worth stressing that we have no guarantee that our improved BP algorithm finds \emph{all} the BP fixed points. Moreover, it was demonstrated that 
while the BP fixed points are Bethe free energy minima, the converse needs not to be true \cite{Heskes2003}.
That said, the high probability with which we find the GS for each sample makes us confident that with this algorithm we are
finding most of the low free-energy BP fixed points.

\begin{figure}[t]
\begin{center}
\includegraphics[height=6cm] {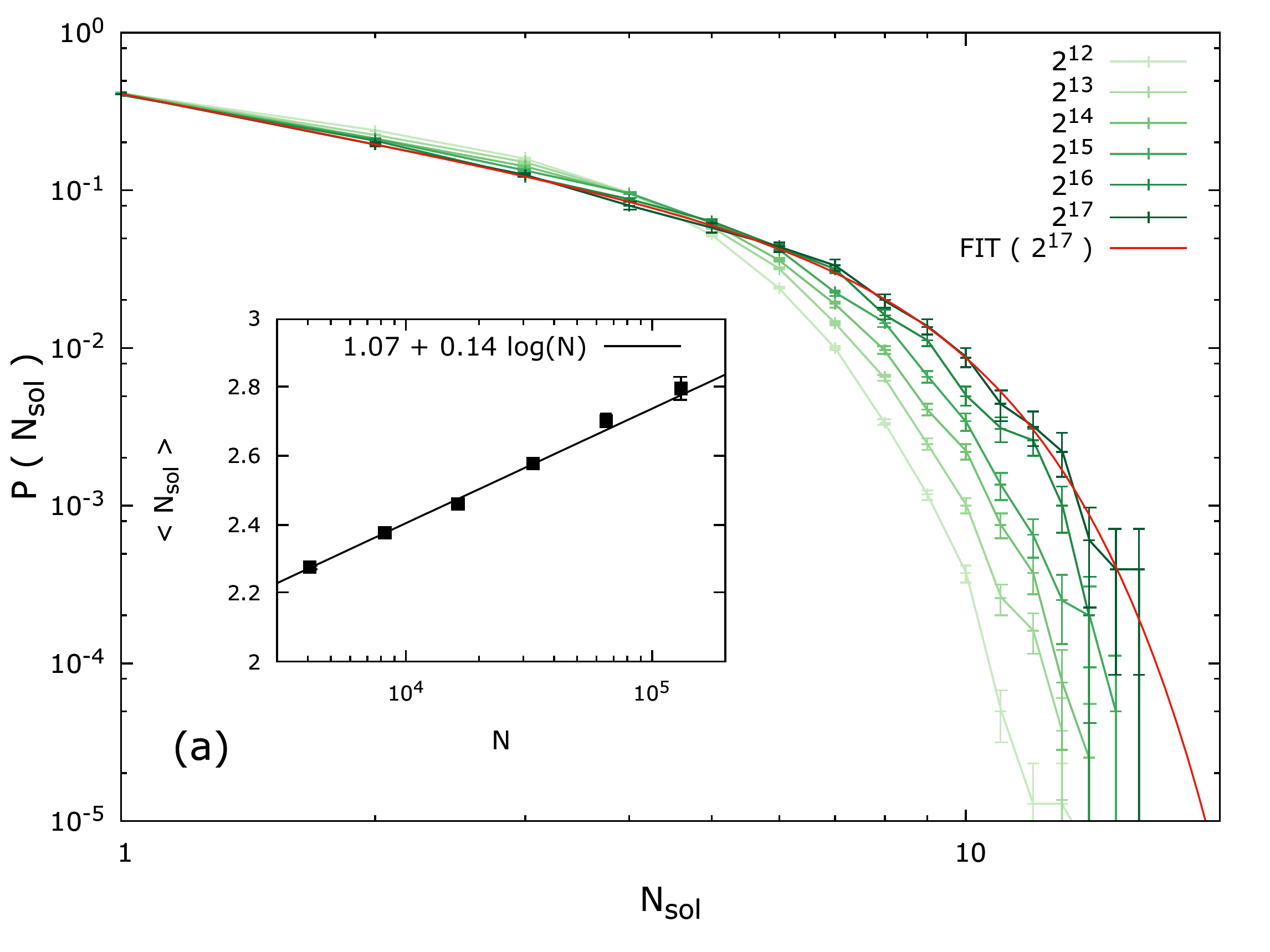}
\includegraphics[height=6cm] {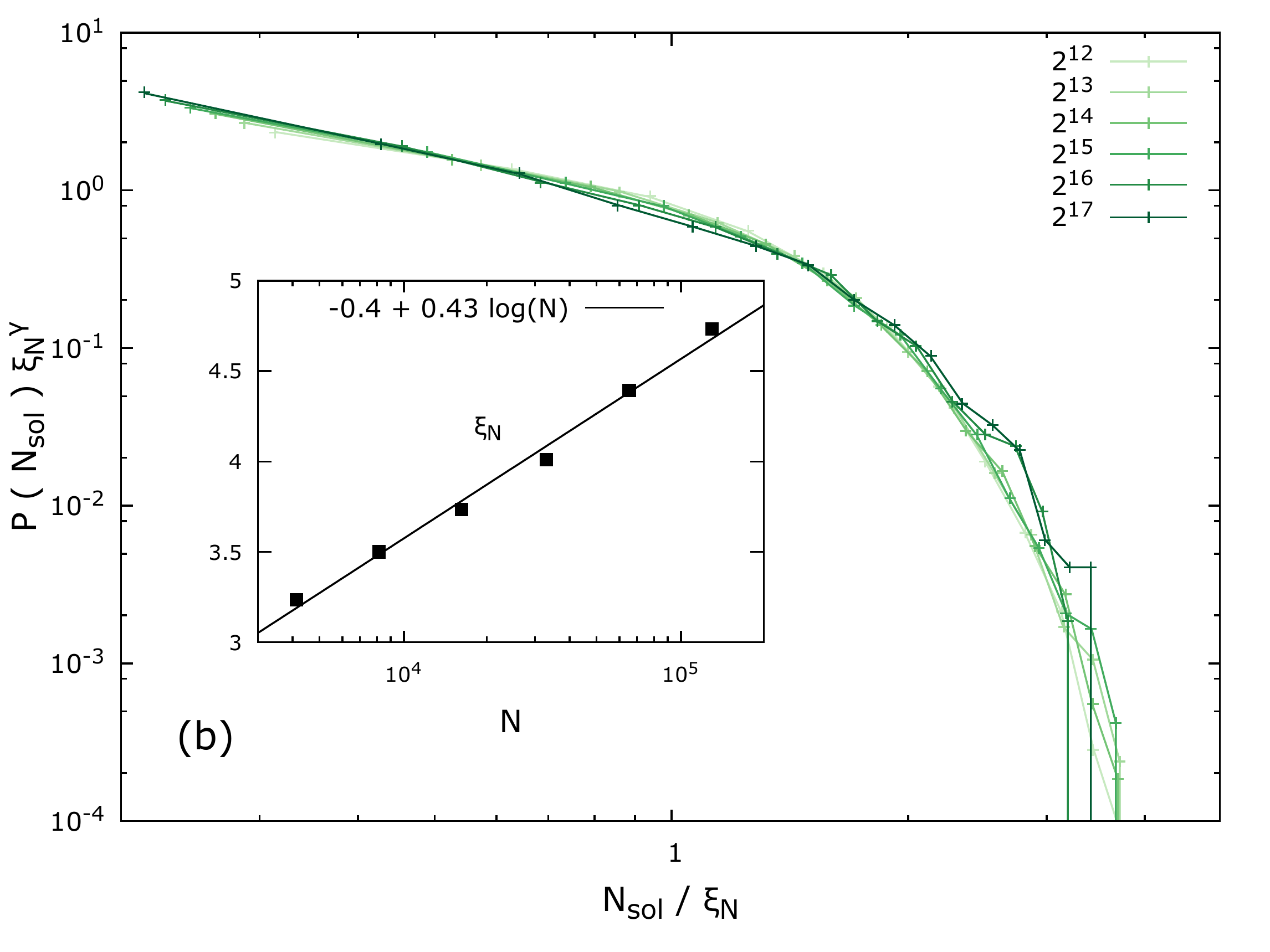}
\end{center}
\caption{(a) Probability distribution of $N_\text{sol}$ at the critical point $J_c$ for several system sizes.
The red full curve is the fit $N_\text{sol}^{-\gamma} \exp(-(N_\text{sol}/\xi_N)^\delta)$ to $N=2^{17}$ data.
The growth of the mean value $\langle N_\text{sol} \rangle$ is shown in the inset, together with the best fit $a + b \log(N)$ where $a=1.07\pm0.04$ and $b=0.145\pm0.004$. 
(b) A rather good collapse of the data shown in the left panel can be achieved by plotting $P(N_\text{sol}) \xi(N)^\gamma$ versus $N_\text{sol}/\xi(N)$, where $\xi(N) = c + d \log(N)$ with $c=-0.4\pm0.2$ and $d=0.43\pm0.02$ is the best linear fit to the $\xi_N$ data shown in the inset.}
\label{fig:nSolRRG_Jc}
\end{figure} 

Let us discuss now how the number of Bethe states grows with $N$ in the critical region. Thanks to the good scaling in the critical region (see inset of Fig.~\ref{fig:nSolRRG}), we can concentrate on studying such a growth exactly at the critical value $J_c$. The inset of the left panel in Fig.~\ref{fig:nSolRRG_Jc} shows the mean number of BP solutions found at $J_c$ as a function of the system size. A linear fit in $\log(N)$ interpolates the data perfectly (fitting with a power law, one gets a very small exponent, usually not larger than 0.1, concluding that the power law fit is not very reliable). The main panels in Fig.~\ref{fig:nSolRRG_Jc} show the entire probability distributions of $N_\text{sol}$ for several values of the system size $N$. The data can be very well interpolated via the function $f(x)\propto x^{-\gamma} \exp(-(x/\xi)^\delta)$ (the red full line in the left panel is the best fit to the $N=2^{17}$ data). We have fixed $\delta=5/2$ as its best fitting value is always very close to 2.5 for all $N$ values. The values of the cutoff $\xi_N$ are shown in the inset of the right panel and can be well fitted by the linear function $\xi(N) = c + d \log(N)$ with $c=-0.4\pm0.2$ and $d=0.43\pm0.02$.
The best fit exponent $\gamma$ increases with $N$ and goes above 1 for the largest $N$ values; however a precise extrapolation to the $N\to\infty$ is difficult. We have estimated the value $\gamma\sim 3/2$ (used to rescale data in the right panel of Fig.~\ref{fig:nSolRRG_Jc} by studying $\langle N_\text{sol} \rangle \propto \xi_N^{2-\gamma}$ versus $\langle N_\text{sol}^2 \rangle / \langle N_\text{sol} \rangle \propto \xi_N$.

\section{Conclusions}

We have presented an improved BP algorithm that is able to find many stable solutions to BP equations (i.e.\ free-energy minima or Bethe states) in the RFIM defined on a random graph.
While the standard implementation of BP is effective only far from the critical region, close to the critical point
the choice of the initial condition plays a crucial role: indeed the number of free-energy minima grows and their basins of attraction shrink, such that a randomly chosen initial condition is unlikely to find the lowest free-energy minima. To partially overcome this problem, we have proposed a new way of recursively initialize BP with a proper interpolation of BP fixed points already found. This algorithm returns the ground state with high probability, together with many other BP fixed points, corresponding to metastable states of low free-energy.

A careful analysis of the number of BP fixed points in the critical region reveals a slow divergence of its mean value $\langle N_\text{sol} \rangle \sim \log(N)$ and a probability distribution decaying, in the large $N$ limit, roughly as $N_\text{sol}^{-3/2}$.
The existence of a diverging number of states is a key feature of disordered systems (e.g.\ is at the heart of the replica symmetry breaking theory \cite{Mezard1987}).
However, to the best of our knowledge, this is the first case where many different Bethe states are explicitly found in a model.
It is still unclear what are the physical consequences of the existence of a diverging number of Bethe states in the critical region of the RFIM; although the thermodynamics of the model can be solved within a replica symmetric ansatz \cite{Krzakala2010}, one is tempting to interpret the occurrence a diverging number of Bethe states as a reminiscence of a replica symmetry breaking phase.

It is important to remark that the Bethe states found by our improved BP algorithm are much more stable than typical minima of the energy potential function; the latter are called one-spin-flip stable configurations and are found in a much broader range \cite{Detcheverry2005, Rosinberg2009}, however their relevance for the out-of-equilibrium dynamics and in general for the determination of physical properties of the the model is unclear.
On the contrary, it has been shown \cite{Weiss2001} that BP fixed points at zero temperature correspond to configurations that are stable under the flip of spins in any subset of vertices forming a subgraph with at most one loop.
This property of BP fixed points is known as maximal neighborhood stability property and makes them much better candidates to describe also the physics at finite temperature and the behavior of thermal algorithm, that may get trapped in these Bethe states even if their evolution is stochastic.

It has been recently shown rigorously \cite{Coja2017} that the Gibbs measure of any random graphical models can be decompose into a moderate number, e.g.\ $O(\log(N))$, of Bethe states, and that the probability marginals in these Bethe states can be obtained from the corresponding BP fixed point. This result implies that, being able to find the relevant BP fixed points, one could in principle compute exactly any marginal in a random graphical model. Our improved BP algorithm allows one to perform this program for the case of the RFIM on a random graph.

In the light of the present results, supporting the existence of a large number of BP fixed points in the RFIM on a random graph, we believe it would be useful to reconsider the proofs that were derived assuming the existence of a unique BP fixed point \cite{Coja2016}.

Let us conclude with a remark on the optimization problem of finding the lowest energy configuration.
For the case of the RFIM on a random graph, our improved BP algorithm finds almost certainly the ground state in a time which is competitive with algorithms, like min-cut, that provably return the exact ground state.
However there are applications where having at hand many low energy configurations, as those retuned by our improved BP algorithm, allows for a much better final choice \cite{Tappen2003}.
Moreover this improved algorithm, finding several different low-energy configurations, allows one to study the fundamental excitations of the system, without the use of methods, like e.g.\ the $\epsilon$-coupling algorithm \cite{Zumsande2009}, that require to modify the Hamiltonian with somehow ad-hoc and not fully justified perturbations.
Finally the algorithm presented here is quite robust with respect to small changes in the Hamiltonian (e.g.\ the introduction of a small fraction of negative coupling), while the min-cut algorithm can not be used as soon as any small amount of frustration is introduced in the interaction couplings.

\begin{acknowledgments}
We thank Giorgio Parisi for useful discussions. This research has been supported by the European Research Council (ERC) under the European Unions Horizon 2020 research and innovation programme (grant agreement No [694925]).
\end{acknowledgments}

\bibliographystyle{unsrt}
%\bibliography{T0_RFIM}
\bibliography{mybib}

\end{document}